\documentclass[sigconf,nonacm]{acmart}

\usepackage{amsmath}
\usepackage{booktabs}
\usepackage{multirow}
\usepackage{graphicx}
\usepackage{comment}

\begin{document}

\title{Deployment Risk Assessment Using Diff-Aware Features: A Case Study at Prime Video}

\author{Mayur Kurup}
\authornote{Both authors contributed equally to this work}
\affiliation{%
  \institution{Amazon.com}
  \city{Austin}
  \state{Texas}
  \country{USA}
}

\author{Hyunjae Suh}
\authornotemark[1]
\authornote{Work done during Amazon internship}
\affiliation{%
  \institution{University of California}
  \city{Irvine}
  \state{California}
  \country{USA}
}

\author{Swathi Vaidyanathan}
\affiliation{%
  \institution{Amazon.com}
  \city{Seattle}
  \state{Washington}
  \country{USA}
}

\author{Pranesh Vyas}
\affiliation{%
  \institution{Amazon.com}
  \city{Austin}
  \state{Texas}
  \country{USA}
}

\author{Srinidhi Madabhushi}
\affiliation{%
  \institution{Amazon.com}
  \city{Seattle}
  \state{Washington}
  \country{USA}
}

\author{Yegor Silyutin}
\affiliation{%
  \institution{Amazon.com}
  \city{Seattle}
  \state{Washington}
  \country{USA}
}

\begin{abstract}
At Amazon Prime Video, we face the critical operational challenge of managing code deployments during live events and rapid feature releases without causing service outages. Current change control approaches use blanket deployment freezes that block all changes regardless of risk, creating significant developer toil. While prior research has explored risky change predictors, these rely on developer-specific metadata or extensive historical data, raising privacy concerns and limiting applicability to new projects. We introduce a framework centered on diff-aware features---characteristics derived directly from code modifications. Our key contribution is the systematic identification of which quantitative metrics (code-level and change-level metrics) and qualitative indicators (coding style violations, change type classification) are necessary for risk prediction. We employ LLMs as multi-language feature extractors, demonstrating their effectiveness for code analysis beyond generation tasks and eliminating the need for language-specific tooling. We evaluated our framework on two datasets: Prime Video's production environment and the public ApacheJIT dataset. Our best-performing model achieves an average recall of \textbf{0.83} and F1 score of \textbf{0.81} across both datasets for detecting risky code changes. Notably, ablation analysis reveals that change-level volume metrics (e.g., lines added/deleted) are noisy predictors, while structural code complexity provides a substantially stronger risk signal. These results demonstrate that thoughtful feature curation enables effective change risk assessment across different programming languages and organizational contexts while avoiding privacy concerns.
\end{abstract}



\maketitle

\section{Introduction}

Live event streaming services such as Prime Video operate under stringent reliability requirements, particularly during large-scale broadcasts like NFL Thursday Night Football (TNF) and Premier League matches, where millions of viewers depend on uninterrupted service. Even a single faulty code change can cause widespread disruptions that are difficult to mitigate during events. To reduce this risk, organizations commonly employ broad deployment freezes during high-traffic windows. While these freezes prevent failures, they delay feature rollouts, create pending change accumulations, and require special handling for urgent updates. As the number and complexity of live events continue to grow, the operational cost of such freezes increases correspondingly, making more precise assessment of individual code-change risk increasingly important.

The central challenge is determining which changes warrant caution without relying on coarse, event-long freezes. Although prior research on change- and defect-prediction provides useful insights~\cite{kamei2012large,6772130}, many approaches depend on developer-specific metadata or substantial historical project information. These dependencies raise privacy considerations, hinder adoption across diverse teams, and limit applicability to new or rapidly evolving codebases. Other models demonstrate effectiveness only within a single dataset, providing limited evidence of generalizability across languages, repositories, or organizational environments.

These limitations motivate an approach using only code modification information. Diff-level characteristics avoid sensitive metadata, apply immediately to new projects including cold-start scenarios such as new code bases, and behave consistently across heterogeneous environments. Such a model would be particularly valuable in contexts like Prime Video's live event operations, where accurate and lightweight risk assessment is needed to make informed deployment decisions during TNF and other high-visibility broadcasts.

In this work, we investigate the predictive utility of diff-aware features for assessing risky code changes. Our intended deployment model is as a \emph{guardrail}: the system flags potentially risky changes for human review rather than autonomously blocking deployments, keeping engineers in the loop for final deployment decisions. This human-in-the-loop design motivates our preference for recall over precision: it is preferable to surface a change for review unnecessarily than to miss a genuinely risky one, leading us to select a threshold that weights recall more (Section~\ref{subsec:classification}). By constructing a comprehensive set of quantitative and qualitative attributes derived directly from code modifications, and evaluating them across both industrial and public datasets, we assess whether code-change information alone can support practical and reliable risk prediction. Our results indicate that a well-curated set of diff-level features captures meaningful risk signals, enabling more targeted and efficient deployment decision-making in environments where reliability and development velocity must be jointly balanced.

Our contributions are: (1) a diff-aware risk assessment framework applicable to cold-start scenarios without developer metadata, (2) LLM-based multi-language feature extraction that replaces brittle language-specific AST parsers, and (3) empirical evidence that LLMs are effective for structured feature extraction but lag behind feature-engineered models for end-to-end risk classification.

The remainder of this paper is organized as follows: Section~2 reviews related work, Section~3 describes our methodology including data collection, feature extraction, and classification, Section~4 presents the classification and ablation results, Section~5 discusses industrial deployment considerations and feature insights, Section~6 addresses threats to validity, and Section~7 concludes with future directions.

\section{Related Work}

Just-in-time (JIT) defect prediction aims to estimate whether a code change is likely to introduce a defect at the moment it is submitted. Kim et al.\ introduced change classification using features derived from change size, diffusion, and file entropy~\cite{kim2008classifying}, establishing the per-commit granularity as a practical setting for early defect detection. Kamei et al.\ conducted large-scale empirical studies demonstrating that inspecting a small portion of commits can prevent a substantial fraction of defect-inducing changes~\cite{kamei2012large}, with change size and complexity metrics (diff-aware features) emerging as the strongest predictors while developer experience (diff-opaque) provided complementary signal. Their cross-project analysis~\cite{kamei2016studying} showed JIT models can transfer across projects under certain conditions, relevant to cold-start scenarios where historical data is unavailable. Deep learning approaches have been applied to learn semantic representations: Hoang et al.\ proposed DeepJIT, jointly encoding commit messages and diffs using CNNs~\cite{hoang2019deepjit}, with ablation studies showing code-level features alone achieved 85--90\% of full model performance, demonstrating that diff-aware features carry substantial predictive power independently. Pre-trained models such as CodeBERT~\cite{feng2020codebert} have been adapted to defect prediction, but these approaches use models as end-to-end classifiers requiring fine-tuning and large corpora. In contrast, we use LLMs strictly as feature extractors---leveraging their multi-language understanding to compute structured metrics, then feeding those into lightweight ML classifiers. This avoids the cost and opacity of end-to-end models while retaining language generalization. Other research investigates finer granularity: Pascarella et al.\ explored file-level JIT prediction~\cite{pascarella2019fine}, and Zeng et al.\ highlighted reproducibility issues in deep JIT approaches~\cite{zeng2021deep}.

Despite extensive work, many models depend on developer-specific metadata, detailed historical traces, or computation-heavy deep learning pipelines. Such dependencies create challenges for industrial adoption, particularly where privacy constraints and rapid onboarding of new projects limit data availability. Our work complements prior research by focusing strictly on \emph{diff-only} features, enabling a metadata-free and lightweight approach suitable for operational settings such as Prime Video's live-event streaming environment. Our findings validate that diff features alone can achieve strong performance (F1: 0.846), directly addressing cold-start scenarios common in industrial settings with frequent repository creation and strict privacy requirements.

\section{Methodology}
\begin{figure}[t]
  \centering
  \includegraphics[width=1\linewidth]{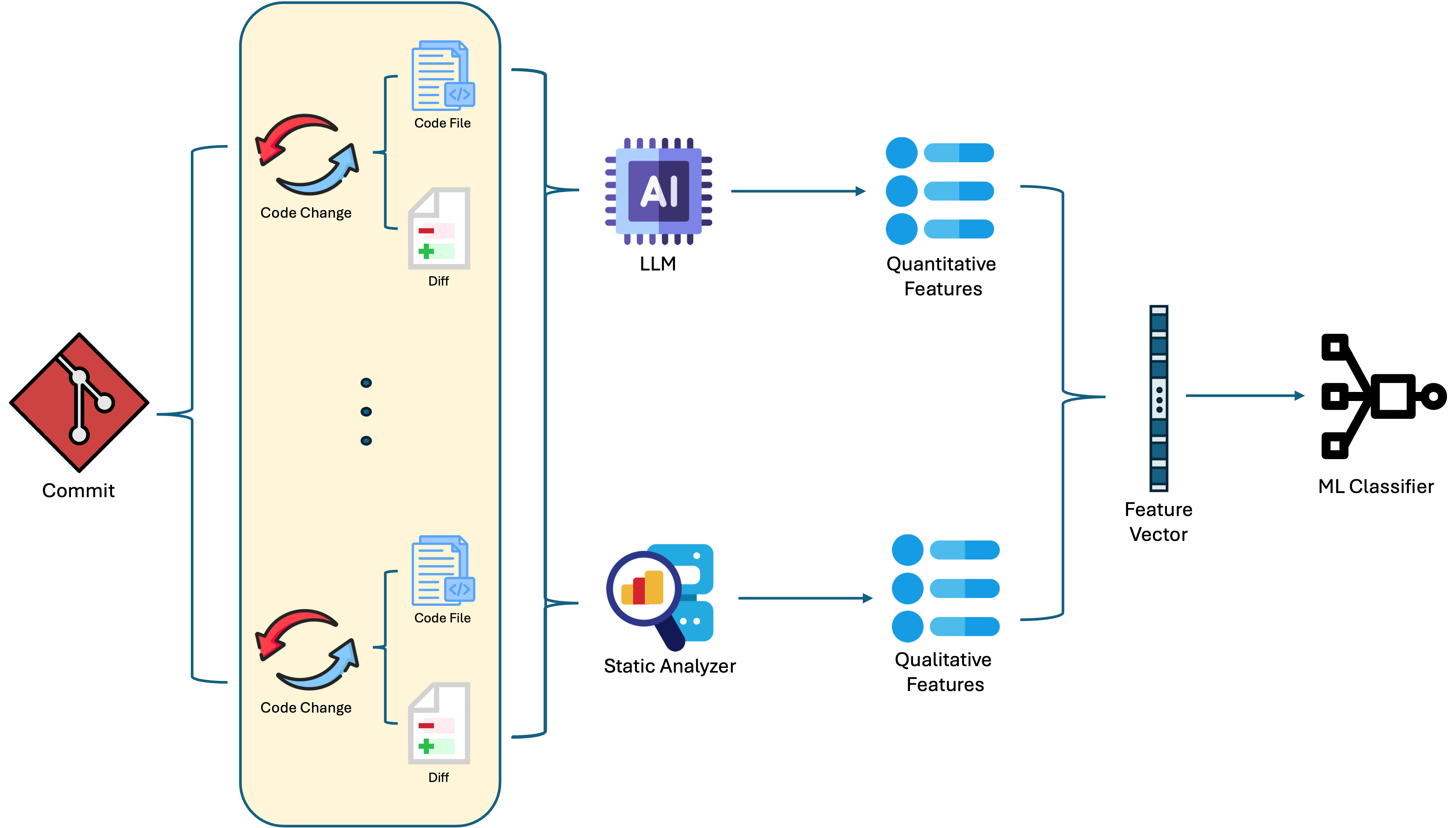}
  \caption{Methodology Overview}
  \label{fig:diagram}
\end{figure}

We describe our methodology, spanning data collection, feature extraction, and classification. Figure~\ref{fig:diagram} presents an overview of the framework that extracts diff-aware features from commits using LLMs for quantitative metrics and static analyzers for qualitative indicators, then applies ML classifiers to predict change risk.

\subsection{Data Collection}

We collected ground truth data by linking Correction of Error (CoE)\cite{aws_coe_2020} reports, a mandatory post-incident review process in which engineers document root causes and link the specific commits responsible, to the corresponding commits that directly triggered the incident relating to the outage. Starting with 377 CoEs flagged as change-induced, we parsed each report to extract the Code Reviews (CRs) identified as root causes. Extracting structured information from CoE reports was challenging due to inconsistent formatting across teams. We used Claude Sonnet 3.7 to parse these reports and extract commit references. Domain experts then manually verified all extracted CoE-to-commit linkages from 2024 to ensure correct attribution, confirming that each linked commit was indeed the root cause identified in the corresponding CoE report. We ranked packages by the number of associated CRs and selected the top 40 packages with the highest CR occurrences, yielding a total of 100 CRs. For each CR, we collected the corresponding commit linked to it. These commits form our ground truth dataset of risky changes.

To focus on widely-used languages in our dataset, we selected Java, Kotlin, and TypeScript for our experiments, which are the predominant languages in the Prime Video codebase. After analyzing our deployment data, we found that approximately 5\% of commits to production result in CoE-flagged incidents. We prioritized labeling precision over dataset size, retaining only commits with high-confidence CoE linkage, and maintained the natural 1:19 risky-to-safe ratio, resulting in 149 risky and 2,831 non-risky commits (2,980 total) that reflect realistic production conditions. For external validation, we used the ApacheJIT dataset~\cite{apachejit2022}, using the most recent three years of data across all 14 Apache projects and sampling 1,115 bug-inducing and 21,185 clean commits (22,300 total) to maintain the 1:19 class ratio.

\subsection{Feature Extraction}

This study leverages diff-aware features extracted from software changes to evaluate change riskiness. We focus on diff-aware features for the following reasons:

\begin{itemize}
\item \textbf{Direct Causality to Change:} Diff-aware features have clear causal connections to potential issues, as changes in code structure, size, and complexity directly impact risk.
\item \textbf{Immediate Applicability:} Diff-aware features apply to new projects or repositories without historical data, supporting first-time contributors and scenarios with limited project history.
\item \textbf{Lower Maintenance:} Diff-aware features eliminate the need to maintain historical data or complex metadata, simplifying implementation and reducing computational overhead.
\end{itemize}

While diff-opaque features (e.g., developer experience, historical defect rates) could potentially enhance performance, they raise privacy concerns and limit generalizability; exploring their integration represents a promising direction for future work. We extracted 28 features categorized into quantitative and qualitative groups. Quantitative features include: (1) \emph{code-level metrics} capturing file structure (lines of code, comments, blank lines, declarations), complexity (cyclomatic complexity, maximum nesting depth), and scope (class, method, and function counts with visibility modifiers); and (2) \emph{change-level metrics} measuring modification extent (added/deleted/modified lines and chunks, churn, files changed). Qualitative features include change type classification (Adaptive, Corrective, Perfective, Preventive) derived from commit messages and coding style violations detected across 13 categories (e.g., naming conventions, code organization, documentation issues).

To extract quantitative features, we employed Claude Sonnet 3.7, which generalizes across multiple languages (Java, Kotlin, TypeScript)
without language-specific reimplementation. We initially developed language-specific scripts using regular expressions and AST patterns, but maintaining separate extractors proved brittle; we therefore adopted LLM-based extraction, retaining the scripts as validation baselines. We validated LLM-extracted features against script-based ground truth (n=60 samples across Java/Kotlin/TypeScript) using two metrics: Pearson correlation~\cite{pearson1895}, which measures linear agreement in magnitude and trend, and Intraclass Correlation Coefficient (ICC)~\cite{shrout1979icc}, which assesses measurement reproducibility. The validation achieved Pearson correlation of 0.730 and ICC of 0.861~\cite{Li2023StaticAnalysisLLMs}.

\begin{table}[hbt]
\centering
\caption{Comparative Analysis Results}
\label{tab:comparative_analysis_results}
\begin{tabular}{lc}
\hline
\textbf{Metric} & \textbf{Average} \\
\hline
Pearson Correlation        & 0.730 \\
Intraclass Correlation (ICC) & 0.861 \\
\hline
\end{tabular}
\end{table}

Table~\ref{tab:comparative_analysis_results} presents the comparison results between the feature values from the script and the LLM. 
The Pearson correlation coefficient and ICC indicate a strong linear relationship between the two sets of values. These results demonstrate that while LLM-extracted values may not be identical in magnitude to script-generated ones, they capture underlying trends and relative patterns with reasonable reliability, making LLMs a viable alternative for feature extraction in change risk assessment.

\begin{table}[h]
\centering
\caption{Style Categories}
\label{tab:style_categories}
\renewcommand{\arraystretch}{1.1}
\small
\begin{tabular}{p{0.28\linewidth} p{0.65\linewidth}}
\hline
\textbf{Category} & \textbf{Description} \\ \hline
Annotations & Annotation usage and placement. \\
Block Checks & Rules for braces and block style. \\
Class Design & Class structure and design rules. \\
Coding & General coding practice checks. \\
Headers & File and license header requirements. \\
Imports & Rules for import order, unused imports, and duplicates. \\
Comments & API documentation format and completeness. \\
Metrics & Complexity and size measurements. \\
Miscellaneous & Other uncategorized checks. \\
Modifiers & Access and modifier order/usage. \\
Conventions & Naming rules for classes, methods, and variables. \\
Regexp & Regex-based custom checks. \\
Size Violations & Limits on line, method, or class length. \\
Whitespace & Spacing, indentation, and line break rules. \\ \hline
\end{tabular}
\end{table}

For extracting the type of change, we adopted categories from previous works~\cite{LevinYehudai2017} (Adaptive, Corrective, Perfective, Preventive). We prompted Claude Sonnet 3.7 to classify the type of change into one of the four categories. To verify the correctness of the LLM's output for this specific feature, we performed stratified sampling across all four categories (n=50) with independent annotation by domain experts.

To extract style violations, we employed static analyzers rather than LLMs, as style violation identification requires consistency and objectivity. We used language-specific analyzers: Checkstyle for Java~\cite{checkstyleCheckstyle1101}, Detekt for Kotlin~\cite{detektHelloFrom}, and ESLint for TypeScript~\cite{typescripteslintTypescripteslint}. Since each analyzer uses different rulesets, we manually mapped all rules to common categories (Table~\ref{tab:style_categories}) based on official documentation, ensuring uniformity across languages. For model input, we counted violation occurrences per category and formatted them as numerical features.

\subsection{Classification}
\label{subsec:classification}
We applied machine learning models to classify change riskiness. For commits with multiple file modifications, we created unified commit-level representations by averaging quantitative features across files and diffs, and averaging violation counts across files for qualitative features.

We applied XGBoost and Random Forest classifiers, evaluated using 10-fold cross-validation within each dataset separately with hyperparameter tuning via grid search. Because our framework is designed as a guardrail that flags changes for human review rather than blocking them autonomously, we favor recall over precision: missing a risky change is costlier than surfacing a safe one for review. To select the classification threshold, we performed a grid search over candidate thresholds $\{0.1, 0.15, 0.2, 0.25, 0.3, 0.35, 0.4, 0.45, 0.5\}$ \emph{within} each cross-validation fold to prevent information leakage, optimizing for F1-score. The threshold of 0.2 consistently maximized F1 across folds: given the 1:19 class imbalance, the default 0.5 threshold yields low recall, and lowering the threshold to 0.2 recovers sufficient true positives to improve F1 despite the precision trade-off. Lower thresholds (e.g., 0.1) yielded marginal recall gains at disproportionate precision cost, while higher thresholds (e.g., 0.3+) sacrificed recall without meaningful precision improvement. We evaluated models using precision, recall, and F1-score. As a baseline, we also performed Logistic regression and zero-shot and few-shot prompting with Claude Sonnet 3.7 and Sonnet 4.5, providing the model with the commit diff, commit description, extracted diff-aware features, and business context for risk classification. We additionally evaluated Claude Sonnet 4.5 under both settings to assess whether newer LLMs narrow the gap with feature-engineered models. The few-shot configuration included three risky and three safe labeled examples, stratified across packages to avoid bias, selected randomly with a fixed seed and excluded from evaluation.

We conducted an ablation study to identify the most influential feature groups using our best-performing model. We tested six configurations: three that excluded one feature group while retaining the other two (Quantitative Code-Level, Quantitative Change-Level, or Qualitative), and three that used only a single feature group to isolate its individual impact.

\section{Results}

\subsection{Classification Results}

We trained and evaluated our models using 10-fold cross-validation on both Prime Video and ApacheJIT datasets, consistent with prior JIT studies~\cite{kamei2012large,hoang2019deepjit}, with results presented in Table~\ref{tab:model_performance_comparison}. XGBoost achieved F1-scores of 0.846 ($\pm$0.070) on Prime Video and 0.771 ($\pm$0.048) on ApacheJIT. On Prime Video, it attained recall of 0.788 with precision of 0.926; on ApacheJIT, recall of 0.875 with precision of 0.689, reflecting different dataset characteristics. The higher variance on Prime Video is expected given its smaller dataset size (2,980 vs.\ 22,300 commits). The Logistic Regression baseline achieved F1-scores of 0.145 on Prime Video and 0.158 on ApacheJIT.

\begin{table}[h]
\centering
\caption{Model Performance Comparison (10-fold CV)}
\label{tab:model_performance_comparison}
\small
\begin{tabular}{lccc|ccc}
\hline
\multirow{2}{*}{\textbf{Model}} & \multicolumn{3}{c}{\textbf{Prime Video}} & \multicolumn{3}{c}{\textbf{ApacheJIT$^\dagger$}} \\
& \textbf{Prec.} & \textbf{Rec.} & \textbf{F1} & \textbf{Prec.} & \textbf{Rec.} & \textbf{F1} \\
\hline
LR (Baseline) & 0.116 & 0.199 & 0.145 & 0.124 & 0.218 & 0.158 \\
XGBoost & \textbf{0.926} & 0.788 & \textbf{0.846} & \textbf{0.689} & 0.875 & \textbf{0.771} \\
RandomForest & 0.811 & \textbf{0.830} & 0.815 & 0.603 & \textbf{0.978} & 0.746 \\
\hline
Sonnet 3.7 (Zero-Shot) & 0.087 & 0.641 & 0.153 & 0.098 & 0.810 & 0.175 \\
Sonnet 3.7 (Few-Shot) & 0.137 & 0.615 & 0.216 & 0.115 & 0.885 & 0.204 \\
Sonnet 4.5 (Zero-Shot) & 0.135 & 0.792 & 0.231 & 0.134 & 0.800 & 0.230 \\
Sonnet 4.5 (Few-Shot) & 0.157 & 0.779 & 0.261 & 0.127 & 0.872 & 0.222 \\
\hline
\end{tabular}
\vspace{2pt}
\par\noindent\scriptsize
$^\dagger$~LLM-based classifiers were evaluated on a stratified subsample of 2,980 ApacheJIT commits (149 buggy, 2,831 clean) matching the Prime Video dataset size to enable cost-effective comparison. ML models were evaluated on the full dataset.
\end{table}

The Random Forest model achieved comparable F1-scores (Prime Video: 0.815, ApacheJIT: 0.746). On Prime Video, it outperformed XGBoost in recall (0.830) but with lower precision (0.811). On ApacheJIT, it achieved exceptional recall of 0.978, though with lower precision (0.603).

LLM-based classification with Claude Sonnet 3.7 achieved F1 of 0.153 (zero-shot) and 0.216 (few-shot) on Prime Video. Upgrading to Claude Sonnet 4.5 improved few-shot performance to 0.157 precision and 0.779 recall (F1: 0.261) on Prime Video and 0.127 precision and 0.872 recall (F1: 0.222) on ApacheJIT. While still below the feature-engineered XGBoost model (F1: 0.846), LLM-based classification shows promise with future steps such as fine-tuning on Prime Video's service dependency graph, incorporating diff-opaque features, and hybrid LLM-ML architectures.

Direct comparison with prior JIT models is limited because most report AUC---which averages over all thresholds and error types---rather than threshold-specific metrics suited to our high-recall guardrail setting, and incorporate developer metadata unavailable in our diff-only approach.

\subsection{Ablation Study}

To understand the contribution of each feature group, we performed an ablation study using XGBoost. We tested six variants: three that excluded one feature group while retaining the others, and three that used only a single feature group.

\begin{table}[htb]
\centering
\small
\caption{Ablation Study Results}
\label{tab:ablation_study_result}
\setlength{\tabcolsep}{4pt}
\begin{tabular}{lccc}
\hline
\textbf{Category} & \textbf{Prec.} & \textbf{Rec.} & \textbf{F1} \\
\hline
Baseline                    & 0.926 & 0.788 & 0.846 \\
Exclude Quant. Code-Level   & 0.873 & 0.736 & 0.781 \\
Exclude Quant. Change-Level & 0.888 & \textbf{0.822} & \textbf{0.848} \\
Exclude Qualitative         & \textbf{0.931} & 0.762 & 0.832 \\
Quant. Code-Level Only      & 0.853 & 0.778 & 0.809 \\
Quant. Change-Level Only    & 0.541 & 0.608 & 0.565 \\
Qualitative Only            & 0.759 & 0.737 & 0.743 \\
\hline
\end{tabular}
\end{table}

As presented in Table~\ref{tab:ablation_study_result}, the full model with all feature groups achieves the best performance. Excluding Quantitative Code-Level features caused the largest performance drop (F1: 0.781 vs.\ baseline 0.846), demonstrating their critical importance. Excluding Quantitative Change-Level features slightly improved performance (F1: 0.848), suggesting these features may introduce noise; we retained them because a marginal gain of 0.002 is unlikely to generalize. Excluding Qualitative features resulted in the highest precision (0.931) but lower recall (0.762) and F1 (0.832), indicating these features help balance the model's predictive capabilities.

Among single-group variants, Quantitative Code-Level Only achieved the best F1 (0.809), confirming code-level metrics as the most influential predictors. Qualitative Only maintained reasonable performance (F1: 0.743), while Quantitative Change-Level Only exhibited severe degradation (F1: 0.565), suggesting these features provide limited value in isolation. These results indicate that structural code complexity, not raw change volume, is the primary driver of risk prediction---a practical insight for teams prioritizing which features to invest in when building change risk systems.

\section{Discussion}

\paragraph{Industrial Deployment and Operational Impact.}
Our approach enables selective deployment decisions at the commit level during live events, potentially reducing freeze scope and accelerating feature velocity. Deployment freezes currently block 100\% of changes regardless of risk, though only approximately 5\% of commits historically cause incidents. On held-out 2024 production data, the model achieved 0.926 precision and 0.788 recall, suggesting it could clear the vast majority of safe changes for deployment while flagging most risky ones for human review---substantially reducing freeze scope without compromising safety. The threshold is tunable: lowering it increases recall further at modest precision cost, acceptable for a system that flags for review rather than auto-blocks. Adoption barriers include CI/CD integration, threshold calibration against business impact, and developer trust through transparent explanations. Validating these estimates through shadow-mode deployment during live events is planned as immediate next work.

\paragraph{Trade-offs and Design Constraints.}
While combining diff-aware features with historical metadata could enhance performance, Prime Video's operational constraints make a diff-only approach optimal: microservice churn makes historical data stale, cross-functional onboarding creates cold-start scenarios, and diff-aware features preserve privacy. Our ablation study (F1: 0.846) suggests diminishing returns from metadata enrichment may not justify maintenance costs.

A key practical benefit of LLM-based feature extraction is eliminating language-specific static analysis tooling. Traditional approaches require separate AST parsers and extraction scripts for each language, each maintained independently as languages evolve. A single LLM extractor generalizes across languages with one prompt template, significantly lowering the cost of onboarding new languages or repositories.

\paragraph{Feature Contribution Insights.}
A counterintuitive finding emerged from our ablation study: quantitative change-level features (lines added/deleted, chunks modified) degraded model performance, suggesting raw change volume is a noisy proxy for risk. Code-level structural complexity and qualitative indicators provide more reliable signals, confirmed by the ablation results (complexity-only F1: 0.809 vs.\ size-only F1: 0.565).

These LLM baselines represent the state of the art for end-to-end agentic risk classification without feature engineering. Performance improved from F1 0.153 to 0.231 (zero-shot) and 0.216 to 0.261 (few-shot) across model generations, demonstrating that newer models reduce the gap but do not close it---structured feature curation remains essential. This gap underscores that decomposing diffs into interpretable quantitative and qualitative dimensions captures risk patterns that end-to-end LLM classification misses, while also producing explainable predictions suitable for developer-facing guardrails.

ApacheJIT's lower F1-score (0.771 vs 0.846) likely stems from dataset characteristics: Prime Video's CoE-based labeling versus ApacheJIT's diverse post-release defect labeling. Our findings extend prior observations that code-level features carry substantial predictive power~\cite{kamei2012large,hoang2019deepjit} to a diff-only standalone model for cold-start scenarios~\cite{fukushima2014cross,liu2025human}.

\section{Threats to Validity}

While our models demonstrate strong performance across both industrial (Prime Video) and open-source (ApacheJIT) datasets, results may be influenced by dataset-specific characteristics.

\paragraph{Construct Validity.} Prime Video uses CoE reports while ApacheJIT uses post-release defects as ground truth, representing different operationalizations of ``risky change.'' The Prime Video dataset is relatively small (149 risky commits) because not all faulty changes result in CoE reports, and high-confidence ground truth filtering further reduces the set. Domain experts manually verified all CoE-to-commit linkages from 2024, confirming correct attribution for all reviewed cases.

\paragraph{Internal Validity.} Our feature set may not capture all aspects of a software change. LLM extraction noise (ICC: 0.861) is absorbed during model fitting since it applies identically at train and inference time; cross-validated F1 inherently reflects this. We evaluated LLM-based classification under zero-shot and few-shot settings with two model generations; few-shot yielded modest improvement, with larger gains on newer models (Sonnet 4.5), though fine-tuning could further narrow the gap. We use 10-fold cross-validation rather than temporal splits, consistent with prior JIT work~\cite{kamei2012large,pascarella2019fine}; temporal evaluation is future work.

\paragraph{External Validity.} The language distribution (Java, Kotlin, TypeScript) may limit feature robustness for underrepresented languages. We mitigated this by evaluating across two distinct datasets with different languages, organizational contexts, and labeling practices. Our features remain applicable to AI-generated code, as structural complexity and style violations characterize the change itself regardless of authorship.

\section{Conclusion and Future Work}

We introduced a diff-aware framework for change risk assessment that derives features directly from code modifications, avoiding dependencies on developer metadata or historical data. On Prime Video and ApacheJIT, XGBoost achieved F1-scores of 0.846 and 0.771, with code-level features proving most influential. Future work includes: (1) shadow-mode deployment during live events to validate predictions against real outcomes, followed by A/B testing to measure operational impact, (2) hybrid models with organizational metadata, (3) semantic change analysis, and (4) fine-tuned LLMs, as zero-shot and few-shot prompting showed limited effectiveness, though Sonnet 4.5 improved few-shot F1 by 21\% over Sonnet 3.7, suggesting LLM-based classification is a viable direction with further investment.

\section{Data Availability Statement}
The primary dataset in this study is derived from Amazon's proprietary production systems and cannot be released due to the confidentiality of internal source code, deployment records, and operational data. To support external validation, we additionally evaluate on the publicly available ApacheJIT dataset~\cite{apachejit2022}. All diff-aware features are defined in full in Section~3, enabling reproduction of our feature-extraction pipeline on other codebases.

\bibliographystyle{ACM-Reference-Format}
\bibliography{references}

@article{kim2008classifying,
title={Classifying software changes: Clean or buggy?},
author={Kim, Sunghun and Whitehead, E James and Zhang, Yi},
journal={IEEE Transactions on software engineering},
volume={34},
number={2},
pages={181--196},
year={2008},
publisher={IEEE},
doi={10.1109/TSE.2007.70773}
}

@article{kamei2012large,
title={A large-scale empirical study of just-in-time quality assurance},
author={Kamei, Yasutaka and Shihab, Emad and Adams, Bram and Hassan, Ahmed E and Mockus, Audris and Sinha, Anand and Ubayashi, Naoyasu},
journal={IEEE Transactions on Software Engineering},
volume={39},
number={6},
pages={757--773},
year={2012},
publisher={IEEE},
doi={10.1109/TSE.2012.70}
}

@article{kamei2016studying,
title={Studying just-in-time defect prediction using cross-project models},
author={Kamei, Yasutaka and Fukushima, Takafumi and McIntosh, Shane and Yamashita, Kazuhiro and Ubayashi, Naoyasu and Hassan, Ahmed E},
journal={Empirical Software Engineering},
volume={21},
number={5},
pages={2072--2106},
year={2016},
publisher={Springer},
doi={10.1007/s10664-015-9400-x}
}

@inproceedings{hoang2019deepjit,
title={Deepjit: an end-to-end deep learning framework for just-in-time defect prediction},
author={Hoang, Thong and Dam, Hoa Khanh and Kamei, Yasutaka and Lo, David and Ubayashi, Naoyasu},
booktitle={2019 IEEE/ACM 16th International Conference on Mining Software Repositories (MSR)},
pages={34--45},
year={2019},
organization={IEEE},
doi={10.1109/MSR.2019.00016}
}

@article{feng2020codebert,
title={Codebert: A pre-trained model for programming and natural languages},
author={Feng, Zhangyin and Guo, Daya and Tang, Duyu and Duan, Nan and Feng, Xiaocheng and Gong, Ming and Shou, Linjun and Qin, Bing and Liu, Ting and Jiang, Daxin and others},
journal={arXiv preprint arXiv:2002.08155},
year={2020},
doi={10.18653/v1/2020.findings-emnlp.139}
}

@article{pascarella2019fine,
title={Fine-grained just-in-time defect prediction},
author={Pascarella, Luca and Palomba, Fabio and Bacchelli, Alberto},
journal={Journal of Systems and Software},
volume={150},
pages={22--36},
year={2019},
publisher={Elsevier},
doi={10.1016/j.jss.2018.12.001}
}

@inproceedings{zeng2021deep,
title={Deep just-in-time defect prediction: how far are we?},
author={Zeng, Zhengran and Zhang, Yuqun and Zhang, Haotian and Zhang, Lingming},
booktitle={Proceedings of the 30th ACM SIGSOFT international symposium on software testing and analysis},
pages={427--438},
year={2021},
doi={10.1145/3460319.3464819}
}

@ARTICLE{6772130,
author={Mockus, Audris and Weiss, David M.},
journal={Bell Labs Technical Journal},
title={Predicting risk of software changes},
year={2000},
volume={5},
number={2},
pages={169-180},
doi={10.1002/bltj.2229}}

@misc{checkstyleCheckstyle1101,
  author = {{Checkstyle}},
  title = {Checkstyle -- A Static Analysis Tool for {Java}},
  howpublished = {\url{https://checkstyle.org/}},
  year = {2025},
  note = {Accessed: 2025-09-05}
}

@misc{detektHelloFrom,
  author = {{Detekt}},
  title = {Detekt -- A Static Code Analysis Tool for {Kotlin}},
  howpublished = {\url{https://detekt.dev/}},
  year = {2025},
  note = {Accessed: 2025-09-05}
}

@misc{typescripteslintTypescripteslint,
  author = {{typescript-eslint}},
  title = {typescript-eslint -- {ESLint} and {Prettier} Support for {TypeScript}},
  howpublished = {\url{https://typescript-eslint.io/}},
  year = {2025},
  note = {Accessed: 2025-09-05}
}

@online{aws_coe_2020,
  author       = {Amazon Web Services, Inc.},
  title        = {Why you should develop a correction of error ({COE})},
  year         = {2022},
  url          = {https://aws.amazon.com/blogs/mt/why-you-should-develop-a-correction-of-error-coe},
  note         = {Accessed: 2026-07-01}
}

@inproceedings{Li2023StaticAnalysisLLMs,
author    = {Haonan Li and Yu Hao and Yizhuo Zhai and Zhiyun Qian},
title     = {Assisting Static Analysis with Large Language Models: A ChatGPT Experiment},
booktitle = {Proceedings of the 31st ACM Joint Meeting on European Software Engineering Conference and Symposium on the Foundations of Software Engineering (ESEC/FSE '23)},
year      = {2023},
publisher = {ACM},
address   = {New York, NY, USA},
url       = {https://www.cs.ucr.edu/~zhiyunq/pub/fseivr23_static_chatgpt.pdf},
doi       = {10.1145/3611643.3613078}
}

@inproceedings{LevinYehudai2017,
title = {Boosting Automatic Commit Classification Into Maintenance Activities By Utilizing Source Code Changes},
author = {Levin, Stanislav and Yehudai, Amiram},
booktitle = {13th International Conference on Predictive Models and Data Analytics in Software Engineering (PROMISE)},
year = {2017},
pages = {97--106},
doi = {10.1145/3127005.3127016}
}

@dataset{apachejit2022,
author       = {Keshavarz, Hossein and Nagappan, Meiyappan},
title        = {ApacheJIT: A Large Dataset for Just-In-Time Defect Prediction},
year         = {2022},
version      = {v1},
publisher    = {Zenodo},
doi          = {10.5281/zenodo.5907002},
url          = {https://zenodo.org/record/5907002}
}

@inproceedings{fukushima2014cross,
title={An empirical study of just-in-time defect prediction using cross-project models},
author={Fukushima, Takafumi and Kamei, Yasutaka and McIntosh, Shane and Yamashita, Kazuhiro and Ubayashi, Naoyasu},
booktitle={Proceedings of the 11th Working Conference on Mining Software Repositories (MSR)},
pages={172--181},
year={2014},
organization={ACM},
doi={10.1145/2597073.2597075}
}

@article{liu2025human,
title={Human-in-the-loop online just-in-time software defect prediction: What have we achieved and what do we still miss?},
author={Liu, Xutong and Yan, Zhiyu and Chen, Zhenyue and Xu, Peng and Yang, Yibiao},
journal={Science of Computer Programming},
year={2025},
publisher={Elsevier},
doi={10.1016/j.scico.2025.103296}
}

@article{pearson1895,
title={Note on regression and inheritance in the case of two parents},
author={Pearson, Karl},
journal={Proceedings of the Royal Society of London},
volume={58},
pages={240--242},
year={1895},
doi={10.1098/rspl.1895.0041}
}

@article{shrout1979icc,
title={Intraclass correlations: Uses in assessing rater reliability.},
author={Shrout, Patrick E and Fleiss, Joseph L},
journal={Psychological Bulletin},
volume={86},
number={2},
pages={420--428},
year={1979},
publisher={American Psychological Association},
doi={10.1037/0033-2909.86.2.420}
}

\end{document}